\documentclass[twoside,final]{pro12}
\usepackage[ansinew]{inputenc}
\usepackage{amsmath}
\usepackage[dvips]{graphicx} 
\graphicspath{{./figures/}}

\head{F. Breitling, G. Mann, C. Vocks} {Propagation of energetic electrons
  and type III radio emission} \label{Barrowpaper}
\begin{document}

\title{Propagation of energetic electrons from the corona into interplanetary
  space and type III radio emission} \author{F. Breitling\adress{Astrophysikalisches Institut Potsdam, An der Sternwarte 16,
    14482 Potsdam, Germany}, G. Mann,
  C.Vocks$\,$}


\maketitle \thispagestyle{empty}

\begin{abstract}
During solar flares a large amount of electrons with energies greater than 20
keV is generated with a production rate of typically $10^{36}$ s$^{-1}$. A
part of them is able to propagate along open magnetic field lines through the
corona into interplanetary space. During their travel they emit radio
radiation which is observed as type III radio bursts in the frequency range
from 100 MHz down to 10 kHz by the WAVES radio spectrometer aboard the
spacecraft WIND, for instance. From the drift rates of these bursts in dynamic
radio spectra the radial propagation velocity $V_r$ of the type III burst
exciting electrons is derived by employing a newly developed density model of
the heliosphere. Calculations show that the radio radiation is emitted by
electrons with different $V_r$ and therefore by different electrons of the
initially produced electron distribution.
\end{abstract}

\section{Introduction}
\begin{figure}[htb] \centering
  \includegraphics[trim=1mm 1mm 1mm 5mm,clip,width=.85\textwidth]{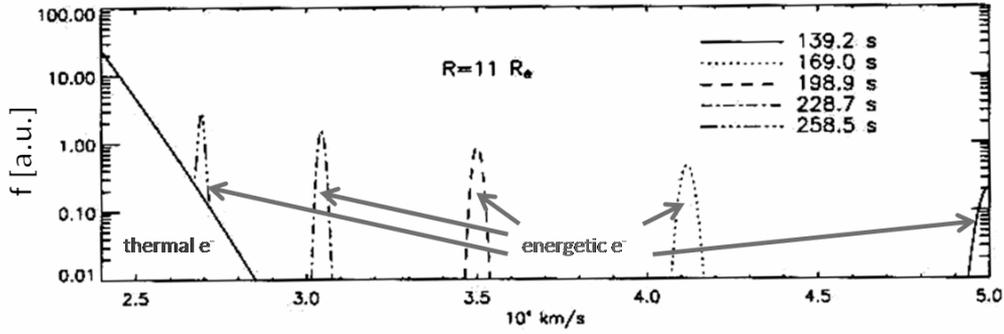}
  \caption{Velocity distribution $f(v)$ of thermal and non-thermal
    energetic electrons in a plasma which cause a ``bump on tail''
    instability. Courtesy by Estel [1999].}
  \label{fig:Estel}
\end{figure}

During solar flares a large amount of energetic non-thermal electrons, i.e.
with energies greater than 20 keV is generated with a production rate of
typically $10^{36}$ s$^{-1}$ (Warmuth et al. [2009]). About 0.1 to
  1\% of them (Lin \& Hudson [1971]) are able to propagate along open
magnetic field lines through the corona into interplanetary space. Signatures
of these electrons can be observed as so called type III bursts in dynamic
radio spectra (Suzuki \& Dulk [1985], Gurnett [1995]). At a greater distance,
the velocity distribution of the electrons develops a bump as shown in Fig.
\ref{fig:Estel}, since slower electrons have not yet arrived and the number of
faster electrons decreases with energy. For comparison Fig. \ref{fig:Estel}
also shows the distribution of thermal electrons in the corona, i.e. the
energetic tail. Non-thermal electron bumps cause so called ``bump on tail''
instabilities which excite Langmuir waves, plasma emission and eventually
radio radiation (Krall \& Trivelpiece [1973]).

\begin{figure}[b!] \centering
  \begin{minipage}[c]{0.6\linewidth}
    \includegraphics[trim=1mm 20mm 10mm
      10mm,clip,width=\textwidth]{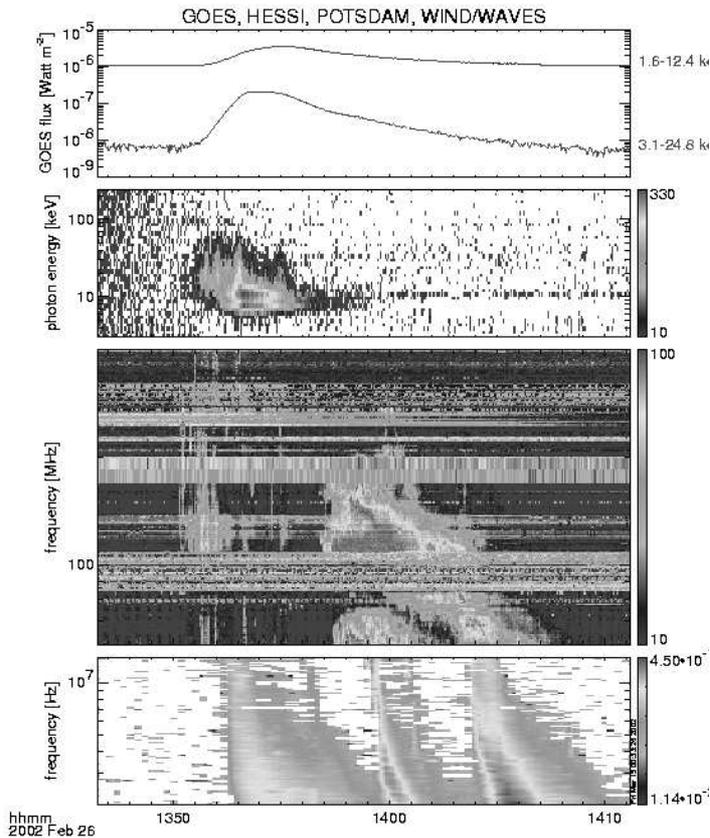}
  \end{minipage} \hfill
  \begin{minipage}[c]{0.32\linewidth}
    \caption{Dynamic spectra of one type II and at least three type III
      radio bursts recorded on 2002 Feb 26 by the X-ray satellites GOES and
      RHESSI, the Observatory for Solar Radio Astronomy (OSRA) and the WIND
      satellite in subsequent order from top to bottom. The type III burst
        are identified by their steep edge and the frequency drift of the
        radio emission from higher to lower frequencies best visible in the
        WIND spectrum.}
    \label{fig:spectrum}
  \end{minipage}
\end{figure}

Fig. \ref{fig:spectrum} shows the typical spectrum of one type II and at
  least three type III bursts. In the upper two plots one sees the energetic
radiation recorded by the X-ray satellites GOES and RHESSI (Lin et al.
  [2002]). In the lower two plots one sees the dynamic radio spectra recorded
by the Observatory for Solar Radio Astronomy (OSRA) of the Astrophysical
Institut Potsdam (Mann et al. [1992]) and the WIND satellite (Bougeret
[1995]). The steep edge and the frequency drift of the radio emission
from higher to lower frequencies is a prominent feature of type III bursts
which is most apparent in the WIND spectrum. Since the radio radiation
originates from plasma oscillations the emission frequency is directly
determined by the plasma density. The density and therefore the plasma
frequency decreases with increasing distance to the Sun due to the Suns
gravitation. Consequently the radio radiation from radially propagating
electrons drifts from higher to lower frequencies. This has been recorded
  for decades in OSRA radio spectra. With the new radio interferometer LOFAR
  (LOw Frequency ARray) radio spectra of higher sensitivity and spectral
  resolution become available in the frequency range from 30 to 240 MHz. For
  example for spectroscopy the effective area increases from about 3 m$^2$ to
  96$\times$ 3 m$^2$ and the spectral resolution improves from 300 to 3 kHz.
  Moreover LOFAR's imaging capabilities allow for the spatial resolution of
  burst propagation. So both instruments provide opportunities to study the
spectra form type III burst in more detail and allow to model the electron
propagation based on the theoretical understanding. Since electron propagation
plays a key roll in space weather its understanding allows for more accurate
space weather predictions which motivates further investigation.

\section{Drift rates and radial propagation velocity}
Plasma emission occurs at the fundamental plasma frequency ($n=1$) and /
  or its first harmonic ($n=2$) (Melrose [1985])

\begin{equation}
  f=\frac{n}{2\pi}\sqrt\frac{e^2N_\mathrm{e}}{\epsilon_0m_\mathrm{e}},
  \label{eqn:f}
\end{equation}

where $\epsilon_0$ is the vacuum permittivity, $e$ and $m_\mathrm{e}$ are the
charge and mass of the electron and $N_\mathrm{e}$ is the plasma's electron
density. Since all other parameters are physical constants, $f$ is a function
of $N_\mathrm{e}$ only.

The drift rate $D_f$ is the frequency change $f$ over time $t$. Using equation
(\ref{eqn:f}) one finds

\begin{eqnarray}
  D_f=\frac{\mathrm{d}f}{\mathrm{d}t}
  =\frac{\mathrm{d}f}{\mathrm{d}N_\mathrm{e}}\frac{\mathrm{d}N_\mathrm{e}}{\mathrm{d}r}\frac{\mathrm{d}r}{\mathrm{d}t}
  =\frac{f}{2N_\mathrm{e}}\frac{\mathrm{d}N_\mathrm{e}}{\mathrm{d}r}\frac{\mathrm{d}r}{\mathrm{d}t},
  \label{eqn:Df}
\end{eqnarray}

where $r$ is the radial distance from the center of the Sun. Consequently the
term $dr/dt$ expresses a radial propagation velocity and will be abbreviated
by $V_r$. Then rewriting equation (\ref{eqn:Df}) yields an expression for the
radial propagation velocity

\begin{eqnarray}
  V_r&=&\frac{2D_f}{f}N_\mathrm{e} \left(\frac{\mathrm{d}N_\mathrm{e}}{\mathrm{d}r}\right)^{-1}.
  \label{eqn:Vr}
\end{eqnarray}

This equation shows, that measurements of the emission frequency and the drift
rate as well as knowledge of the electron density of the interplanetary space
allows to determine the radial propagation velocity $V_r$ of the radio
radiation emitting electrons.

\begin{figure}[tb] \centering
\includegraphics[width=0.6\textwidth]{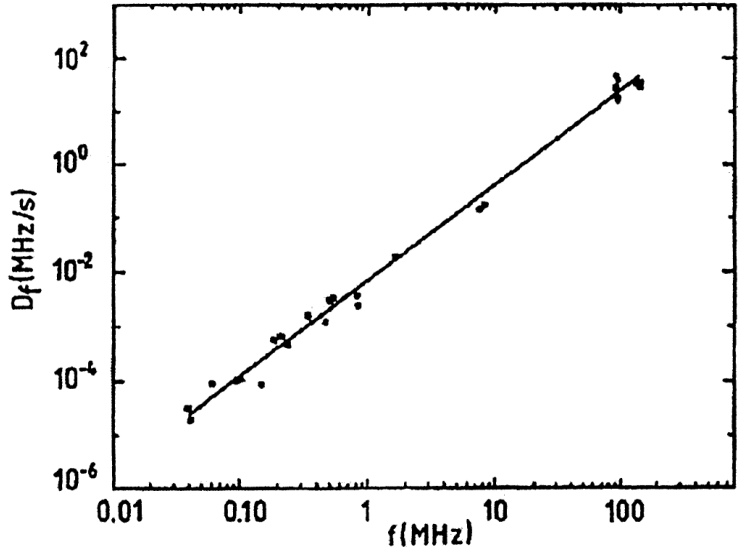}
\caption{Drift rate of a type III radio burst on December 27, 1994 as
  determined by Mann et al. [1999] using measurements from OSRA, WIND and the
  ULYSSES satellite. The fit illustrates the power-law relation of equation
  (\ref{eqn:power-law}).}
  \label{fig:Mann}
\end{figure}

Drift rates can be determined quite accurately from dynamic radio spectra. A
typical example is show in Fig. \ref{fig:Mann} which was taken from Mann
et al. [1999]. It shows that the drift rate, which has been determined at the
leading edge of the type III burst signature, obeys the power-law

\begin{equation}
  D_f=-7.354 \times 10^{-3}
  \mathrm{\frac{MHz}{s}}\left(\frac{f}{\mathrm{MHz}}\right)^{1.76}.
  \label{eqn:power-law}
\end{equation}
However, on small scales and in particular at higher frequencies above 80
  MHz deviations from a power-law are possible.

\section{Density model for the interplanetary space}
According to equation (\ref{eqn:Vr}) the heliospheric electron density is
required for determining the radial propagation velocity $V_r$ from drift rate
measurements. The electron density is directly proportional to the mass
density of the interplanetary plasma $N$ since $N_\mathrm{e}=N/(1.92 \tilde\mu
m_\mathrm{p})$, where $m_\mathrm{p}$ is the proton mass and $\tilde\mu$ is the
mean molecular weight (Priest [1982]). In the inner corona up to a distance of
1.6 solar radii $N$ has been determined from optical measurements (Newkirk
[1961], Koutchmy [1994], Warmuth et al. 2005). From 0.3 to 1 AU $N$ has been
measured by the HELIOS satellite (Schwenn [1990]). Unfortunately it hasn't
been measured in the region from 1.6 solar radii to 0.3 AU. However, one can
apply laws of physics to model the density there. For simplicity we
  neglect local density variations and consider mean radially symmetric models
  only. For example Mann et al. [1999] have applied the Parker [1958] wind
model to data from the Helios satellite to obtain an interplanetary density
model. Ansatz is the equation of motion
\begin{equation}
  mNv\frac{\mathrm{d}v}{\mathrm{d}r}=-\frac{\mathrm{d}p}{\mathrm{d}r}-\frac{GM_\odot mN}{r^2},
  \label{eqn:ofmotion}
\end{equation}
where $m$, $N$ and $v$ are the mean particle mass, density, and velocity in
radial direction $r$, $p$ is the pressure, $G$ is the gravitational constant
and $M_\odot$ is the solar mass.

Here this model is extended by adding heat conduction as well as heating and
pressure by Alfv\'en waves as done by Hackenberg et al. [2000].

Heat conduction is added via the energy equation
\begin{equation}
  vN\left[\frac32k_B\frac{\mathrm{d}T}{\mathrm{d}r}-k_B\frac{T}{n}\frac{\mathrm{d}n}{\mathrm{d}r}\right]=\frac
  1{r^2}\frac {\mathrm{d}}{\mathrm{d}r}\left(r^2\kappa_0\frac{\mathrm{d}T}{\mathrm{d}r}\right)+H,
  \label{eqn:energy}
\end{equation}
where $k_B$ is the Boltzmann constant, $T$ is the temperature, $\kappa_0$ is
the thermal conductivity and $H$ is the heating function by the Alfv\'en
waves.
\begin{equation}
  H(r)=-P_0[v(r)+v_\mathrm{A}(r)] \frac 1{\omega_\mathrm{H}} \frac
  {\frac{v(R_\odot)}{v_\mathrm{A}(R_\odot)}\left(1+\frac{v(R_\odot)}{v_\mathrm{A}(R_\odot)}\right)^2}
  {\frac{v(r)}{v_\mathrm{A}(r)}\left(1+\frac{v(r)}{v_\mathrm{A}(r)}\right)^2}
  \frac{\partial  \omega_\mathrm{H}}{\partial r},
  \label{eqn:heating}
\end{equation}
where $P_0$ is a pressure parameter, $\omega_\mathrm{H}$ is the higher cut-off
frequency, $v_\mathrm{A}$ is the Alfv\'en velocity and $R_\odot$ is the solar
radius.

The Alfv\'en wave pressure $p_\mathrm{A}$ adds the term
$-\mathrm{d}p_\mathrm{A}/\mathrm{d}r$ to the momentum equation
(\ref{eqn:ofmotion}), where
\begin{equation}
  p_\mathrm{A}(r)=\frac12 P_0\log\left(\frac{\omega_\mathrm{H}}{\omega_\mathrm{L}}\right)\frac
  {\frac{v(R_\odot)}{v_\mathrm{A}(R_\odot)}\left(1+\frac{v(R_\odot)}{v_\mathrm{A}(R_\odot)}\right)^2}
  {\frac{v(r)}{v_\mathrm{A}(r)}\left(1+\frac{v(r)}{v_\mathrm{A}(r)}\right)^2},
  \label{eqn:pressure}
\end{equation}
and $\omega_\mathrm{L}$ is the lower cut-off frequency.

\begin{figure}[b] \centering
  \includegraphics[width=0.6\textwidth]{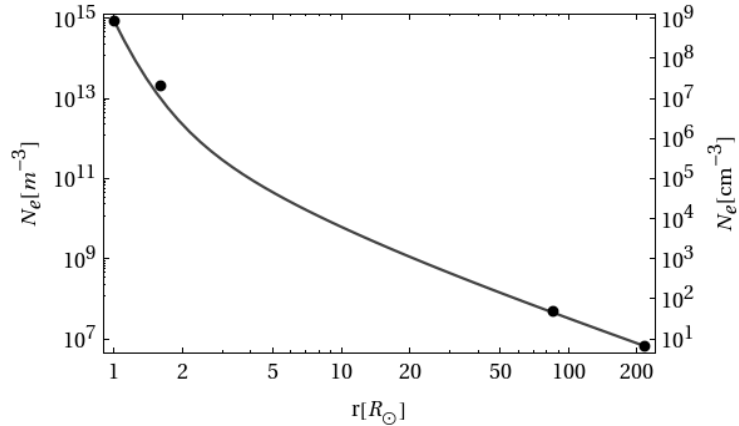}
  \caption{Electron density of the interplanetary space as known from
    measurements (four data points) and the new model described here. For
    numerical values and references see table \ref{table:N}.}
  \label{fig:N}
\end{figure}

\begin{figure}[htb] \centering
  \includegraphics[width=.83\textwidth]{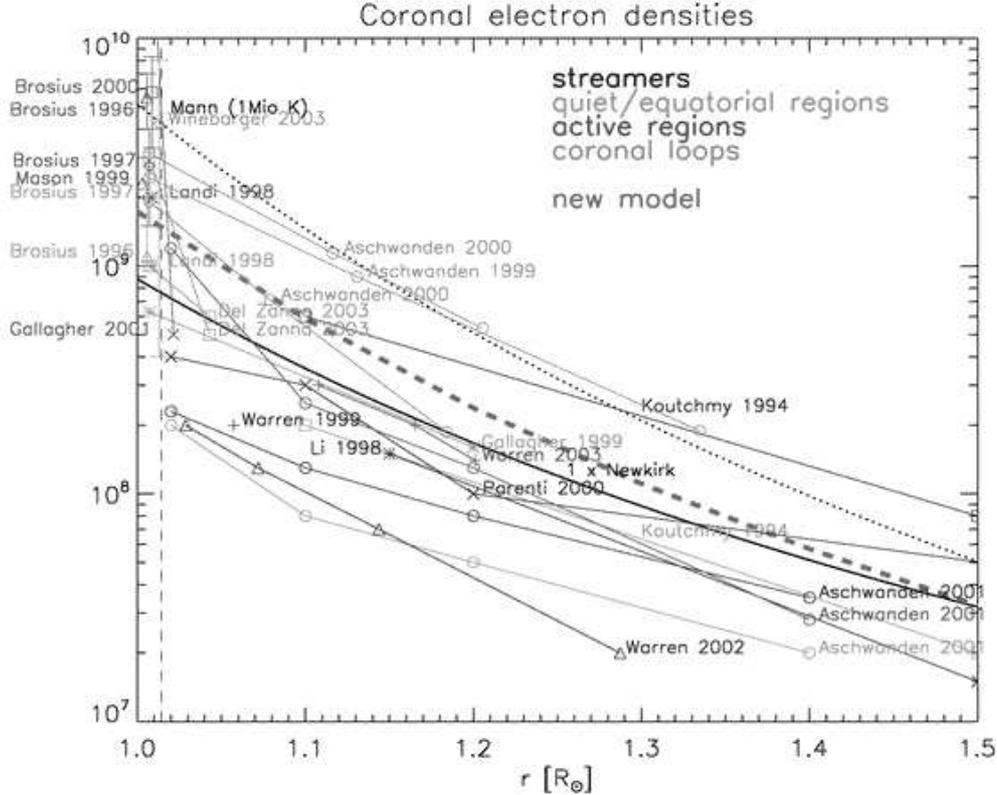}
  \caption{Electron density in inner corona above the transition region
    as collected by Warmuth et al. [2005] from different measurements. The new
    density model described here is shown by the dashed line.}
  \label{fig:Warmuth}
\end{figure}

The result is shown in Fig. \ref{fig:N} and table \ref{table:N}. Table
\ref{table:N} shows good agreement between the new density model and existing
data. Fig. \ref{fig:Warmuth} shows good agreement of this model with density
measurements of the inner corona.

\begin{table}[b]
  \caption{Comparison of parameters from literature with those from the
      new density model. The corresponding references are indicated as
      follows: $^{I}$Schwenn [1990] (HELIOS), $^{II}$Newkirk [1961].}
  \centering
  \medskip
  \begin{tabular}{cccccccc}
    \hline\hline 

    & $v(1\mathrm{AU})$ & $T_{R_\odot}$ & $T_{1\mathrm{AU}}$ &
    $N_\mathrm{e}(R_\odot)$ & $N_\mathrm{e}(1.6R_\odot)$ &
    $N_\mathrm{e}(0.4\mathrm{AU})$ & $N_\mathrm{e}(1\mathrm{AU})$ \\

    & [km/s] & [$10^6$K] & [$10^6$K] & [$10^8$cm$^{-3}$] &
    [$10^7$cm$^{-3}$] & [cm$^{-3}$] & [cm$^{-3}$] \\

    \hline
    Data  & 470$^{I}$ & 1.0-1.5 & 0.14$^{I}$ & 8.78$^{II}$ & 2.10$^{II}$ &
    48.3$^{I}$ & 6.67$^{I}$ \\
    Model & 470       & 1.1     & 0.14       & 8.85        & 1.00        &
    44.6       & 6.67 \\
    \hline
  \end{tabular}
  \label{table:N}
\end{table}

\section{Results}
\begin{figure}[htb] \centering
  \begin{minipage}[t]{0.485\linewidth}
    \includegraphics[width=\linewidth]{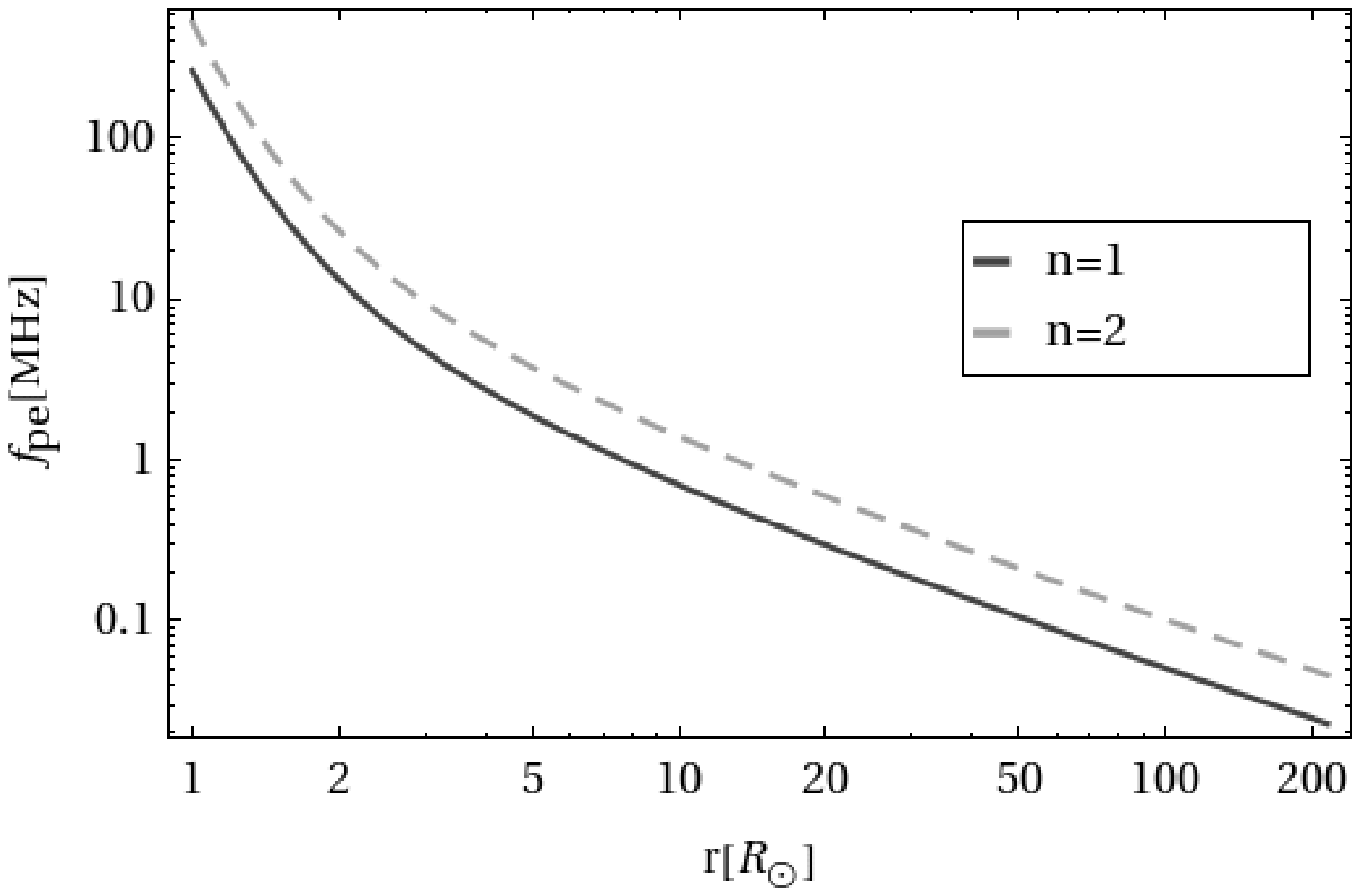}
    \caption{Observation frequency of fundamental (n=1) and first harmonic
      (n=2) plasma emission according to equation (\ref{eqn:f}) and the new
      interplanetary density model of Fig. \ref{fig:N}.}
    \label{fig:fpe}
  \end{minipage}\hfill
  \begin{minipage}[t]{0.485\linewidth}
    \includegraphics[width=\linewidth]{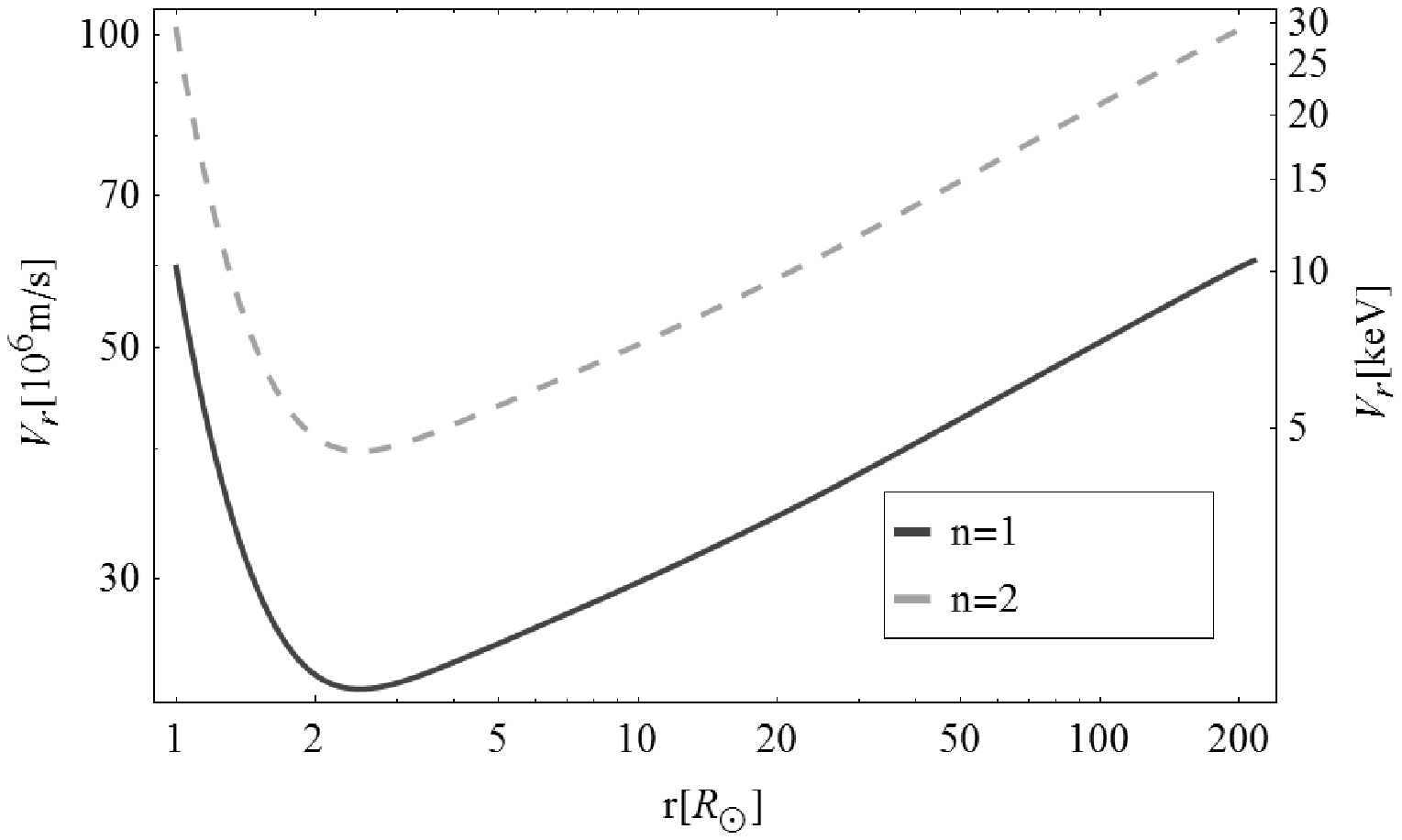}
    \caption{Radial propagation velocity $V_r$ (equation (\ref{eqn:Vr}))
      corresponding to the observation frequencies in Fig. \ref{fig:fpe} and
      the drift rate described by equation (\ref{eqn:power-law}). The
        corresponding electron energies are shown to the right for comparison
        with in-situ measurements.}
    \label{fig:Vr}
  \end{minipage}
\end{figure}

Fig. \ref{fig:fpe} shows the observation frequencies according to the new
density model. One can see that in the region from 1 to 2 solar radii radio
emission is mainly produced in the LOFAR frequency range from 30 to 240 MHz.

Applying the new electron density model to the drift rate measurement of Fig.
\ref{fig:Mann} yields the radial propagation velocity $V_r$ shown in Fig.
\ref{fig:Vr}. While $V_r$ decreases in the inner region from 1 to 2.5 solar
radii it increases beyond. It shows, that the emitting electrons have
different velocities at different distances from the Sun and therefore belong
to different parts of the electron distribution. Future models of the electron
propagation will have to reproduce this feature.

\section{Conclusion}
It was shown how information about the propagation of energetic electrons
emitted by a type III burst can be obtained from dynamic radio spectra. A
consistent electron density model of the plasma in the interplanetary space is
necessary which was developed here. First results from its application yields
a profile for the radial electron propagation velocity $V_r$. It shows that
electrons of different velocities are responsible for the emission of radio
radiation in type III burst and therefor must belong to different parts of the
electron spectrum. It will be left to future work to find models of electron
propagation which can resemble this feature. Next steps will also include
further studies of dynamic radio spectra from solar observations with LOFAR
and OSRA.

\section{Acknowledgement}
The authors want to thank Hakan \"Onel for his support with the computing
software \textit{Wolfram Mathematica 7}. His expertise was very valuable and
highly appreciated.

\section*{References}
\everypar={\hangindent=1truecm \hangafter=1}

Bougeret, J.-L. et al., Waves: The Radio and Plasma Wave Investigation on the
Wind Spacecraft, \textit{Space Science Reviews}, \textbf{71}, 231-263, 1995.

Gurnett, D. A., Heliospheric Radio Emissions, \textit{Space Science Reviews},
\textbf{72}, 243-254, 1995

Estel, C., Ausbreitung energetischer Elektronen von der Sonnenkorona in die
innere Heliosph\"are, \textit{Dissertation}, \textbf{University Potsdam},
1999.

Krall, N., A., Tivelpiece, A., W., Principles of Plasma Physics,
\textit{McGraw-Hill, New York}, 1973

Koutchmy, S., Coronal physics from eclipse observations, \textit{Advances in
  Space Research}, \textbf{14}, (4)29-(4)39, 1994

Hackenberg, P., Marsch, E., Mann, G., On the origin of the fast solar wind in
polar coronal funnels, \textit{Astronomy \& Astrophysics}, \textbf{360},
1139-1147, 2000.

Lin, R. P., Hudson, H. S., 10 - 100 keV electron acceleration and emission
from solar flares, \textit{Solar Physics} \textbf{17}, 412, 1971

Lin, R. P. et al., The Reuven Ramaty High-Energy Solar Spectroscopic Imager
(RHESSI), \textit{Solar Physics}, \textbf{210}, 3-32, 2002

Mann, G., Aurass, H., Voigt, W., Paschke, J., Preliminary observations of
solar type 2 bursts with the new radiospectrograph in Tremsdorf, \textit{in
  Coronal Streamers, Coronal Loops, and Coronal and Solar Wind Composition,
  ESA Special Publication}, \textbf{348}, 129-132, 1992

Mann, G., Jansen, F., MacDowall, R. J., Kaiser, M. L., Sone, R. G., A
heliospheric density model and type III radio bursts, \textit{Astronomy \&
  Astrophysics}, \textbf{348}, 614-620, 1999.

Melrose, D., Plasma emission mechanisms, \textit{in Solar Radio Physics, Ed.
  D. J. McLean \& N. R. Labrum, Cambridge University Press, Cambridge},
177-210, 1985

Newkirk, Jr. G., The Solar Corona in Active Regions and the Thermal Origin of
the Slowly Varying Component of Solar Radio Radiation, \textit{ Astrophysical
  Journal}, \textbf{133}, 1961

Parker, E. N., Dynamics of the interplanetary gas and magnetic fields,
\textit{Astrophysical Journal}, \textbf{128}, 664-676, 1958.

Priest, E. R., Solar magneto-hydrodynamics, \textit{Dordrecht, Holland ;
  Boston : D. Reidel Pub. Co. ; Hingham}, 1982

Schwenn, R., Large-Scale Structure of the Interplanetary Medium, \textit{in
  Schwenn, Rainer, M., E., Physics of the inner heliosphere I, Large-Scale
  Phenomena, Springer-Verlag Berlin Heidelberg New York.}, \textbf{20}, 99
pp., 1990

Suzuki, S., Dulk, G. A., Bursts of Type III and Type V, \textit{in Solar Radio
  Physics, Ed. D. J. McLean \& N. R. Labrum, Cambridge University Press,
  Cambridge}, 289, 1985

Warmuth, A., Mann, G., A model of the Alfven speed in the solar corona,
\textit{Astronomy \& Astrophysics}, \textbf{435}, 1123-1135, 2005.

Warmuth, A., Mann, G., Aurass, H., Modelling shock drift acceleration of
electrons at the reconnection outflow termination shock in solar flares.
Observational constraints and parametric study, \textit{Astronomy \&
  Astrophysics}, \textbf{494}, 677-691, 2009.

\end{document}